\def\bq{\begin{equation}}
\def\eq{\end{equation}}
\def\bqa{\begin{eqnarray}}
\def\eqa{\end{eqnarray}}
\def\bqb{\begin{eqnarray*}}
\def\eqb{\end{eqnarray*}}
\documentstyle[12pt]{article}\textwidth 16cm
\def\pr#1#2#3{ Phys. Rev. ${\bf{#1}}$ (#2) #3}

\def\pl#1#2#3{ Phys. Lett. ${\bf{#1}}$ (#2) #3 }
\def\prep#1#2#3{ Phys. Rep. ${\bf{#1}}$ (#2) #3}
\def\np#1#2#3{ Nucl. Phys. ${\bf{#1}}$ (#2) #3}
\def\zp#1#2#3{ Z. Phys. ${\bf{#1}}$ (#2) #3}

\global\nulldelimiterspace = 0pt

\def\Bsl{\hbox{/\kern-.6700em$B$}} 
\def\Dsl{\hbox{/\kern-.6700em$D$}} 
\def\Wsl{\hbox{/\kern-.6700em$W$}} 

\def\roughly#1{\mathrel{\raise.3ex
    \hbox{$#1$\kern-.75em\lower1ex\hbox{$\sim$}}}}

\def\O{ {\cal O }}
\def\A{ {\cal A }}

\def\mh2{m^2_H}

\def\hbeta{{{\beta} \over 2}}

\begin{document}
\pagenumbering{arabic}
\thispagestyle{empty}
\hspace {-0.8cm} PM/95-40 \\
\hspace {-0.8cm} September 1995\\
\vspace {0.8cm}\\

\begin{center}
{\Large\bf Z' Reservation at LEP2} \\

 \vspace{1.8cm}
{\large  G. Montagna$^a$,
O. Nicrosini$^{b}\footnote{\footnotesize Permanent address: INFN, Sezione
di Pavia, Italy.}$, F. Piccinini$^c$}\\
{\large  F.M. Renard$^d$ and C. Verzegnassi$^e$}
\vspace {1cm}  \\
$^a$Dipartimento di Fisica Nucleare e Teorica, Universit\`a di Pavia\\
and INFN, Sezione di Pavia, Via A.Bassi 6, 27100, Pavia, Italy.\\
\vspace{0.2cm}
$^b$CERN, Theory Division\\
CH 1211 - Geneva 23, Switzerland.\\
\vspace{0.2cm}
$^c$ INFN, Sezione di Pavia, Via A.Bassi 6, 27100, Pavia, Italy.\\
\vspace{0.2cm}
$^d$Physique
Math\'{e}matique et Th\'{e}orique,
CNRS-URA 768,\\
Universit\'{e} de Montpellier II,
 F-34095 Montpellier Cedex 5.\\
\vspace{0.2cm}
$^e$ Dipartimento di Fisica,
Universit\`{a} di Lecce \\
CP193 Via Arnesano, I-73100 Lecce, \\
and INFN, Sezione di Lecce, Italy.\\

\vspace{1.5cm}

{\bf Abstract}
\end{center}
\noindent
We consider the possibility that one extra $Z\equiv Z'$ exists with
arbitrary mass and fermion couplings that do not violate (charged)
lepton universality. We show that, in such a situation, a functional
relationship is generated between the \underline{deviations} from the
SM values of three leptonic observables of two-fermion production at
future $e^+e^-$ colliders that is completely independent of the values
of the $Z'$ mass and couplings. This selects a certain region in the
3-d space of the deviations that is \underline{characteristic} of the
model ($Z'$ "reservation"). As a specific and relevant example, we show
the picture that would emerge at LEP2 under realistic experimental
conditions.

\vspace{1cm}

\setcounter{page}{0}
\def\thefootnote{\arabic{footnote}}
\setcounter{footnote}{0}
\clearpage

The impressive amount of data collected in the last five years at LEP1
and SLC have led to the conclusion that all the theoretical predictions
of the SM (with a possible still conceivably allowed exception for the
$Z$ partial width into $b\bar b$ \cite{LEP})
are in spectacular agreement
with the experimental results, to an accuracy that has reached for some
observables the permille level. This has led to a first "filtering" of
candidate models of new physics, whose effects have been really drastic
only for a limited set of "classical" technicolour schemes \cite{TC},
but rather mild for the majority of competitor proposals
(supersymmetry, anomalous gauge couplings, extra U(1),...). Thus at the
beginning of the second LEP2 phase, the hopes of either producing or
detecting via virtual effects some evidence of new physics are still
well alive for a number of respectable models, with some useful
simplification possibly achieved by taking the LEP1, SLC results into
account.\par
For the specific case of one $Z\equiv Z'$ of the most general
theoretical origin, the relevant information that has been derived is
that the $Z-Z'$ mixing is sufficiently small to be neglected in the
theoretical analyses of two-fermion production at future $e^+e^-$
colliders, which means that only the $Z'$ exchange diagram must be
retained. (This statement might be contradicted for the case of final
$WW$ states produced by longitudinally polarized leptons
\cite{Paver}, that
we shall not consider in this paper).\par

Let us discuss in some more detail this statement. As a matter of fact,
numerical bounds for the mixing angle have been derived for a number of
"canonical" cases of well defined group-theoretical ($E_6$, $LR$
symmetry) origin \cite{Zpmod} and, also, for a $Z'$ of composite models
origin \cite{Zpcomp}. The various results \cite{ZZp}, \cite{Alt} are in
substantial agreement, and suggest a conservative bound of the order of
(at most) one percent. For such a bound, it has already been shown
\cite{Blon}, \cite{Bou} that the mixing effects at future $e^+e^-$ colliders
(for fermion production) are completely irrelevant at the realistic
expected experimental accuracies. In fact, it was shown in
ref.\cite{Blon}
for the specific case of LEP2 that values of the mixing angle much
larger than the final LEP1, SLC bounds (more precisely, values of a few
percent or more) would not be experimentally visible, even for
extremely low ($\simeq 250$~GeV) $Z'$ mass, or, otherwise stated, that
$Z-Z'$ mixing effects can be safely neglected. This conclusion can be
reformulated in a more convenient way, that allows to extend it to the
case of a  $Z'$ with arbitrary fermion couplings, just by noticing
that, in fact, the LEP1, SLC bounds are obtained for several
\underline{products} of the type $ \theta_M g'_{Vf}$, $\theta_M
g'_{Af}$, \cite{ZZp} where $g'_{V,Af}$ are the vector and axial
couplings of the $Z'$ to a generic fermion. If the couplings are
different from the "canonical" ones, the bound for the mixing angle
will change, but those for the products $ \theta_M g'_{V,Af}$
will remain invariant. Since these products are obviously the same that
appear in the various $Z-Z'$ mixing effects at LEP2, the conclusions of
ref.\cite{Blon} remain consequently generally valid. This simplified
argument allows us, in particular, to conclude that we shall be able to
neglect the mixing effects in the purely leptonic processes that we
shall consider, for general universal values of the $Z'$ couplings to
charged leptons.\par
This introductory discussion had a precise motivation. Actually, the
aim of this short paper is that of showing that, from the combined
analysis of leptonic processes at future $e^+e^-$ colliders, it would
be possible to identify the virtual signals of a $Z'$ of the most
general type i.e. with general (but universal) couplings with charged
leptons (no universality assumption on the contrary on the couplings
with the remaining fermions). To prove this statement requires the
discarding of the mixing effects, that would otherwise introduce one
extra unwanted parameter.\par
Considering a most general $Z'$ of the type just illustrated can be
explained, or justified, by two main simple reasons. The first one is
the fact that some of the strong theoretical motivations that supported
"canonical" schemes like e.g. the special group $E_6$ have become
undeniably weak in the last few years. The second one is that a number
of different models with one extra U(1) have meanwhile been proposed,
or have resurrected \cite{Kos}. These facts lead us to the conclusion
that a totally general analysis might be more relevant than a few
specific ones. Obviously, one will be able to recover the "canonical"
results as special cases of our investigation.\par
 In this spirit, we
have started by considering the theoretical expression of the
scattering amplitude of the process $e^+e^-\to l^+l^-$ ($l=e, \mu,
\tau$) at squared c.m. energy $q^2$ in the presence of one $Z'$. At
tree level, this can be written as :

\bq   A^{(0)}_{ll}(q^2) =  A^{(0)\gamma,Z}_{ll}(q^2) +
A^{(0)Z'}_{ll}(q^2)\eq

\noindent
where

\bq    A^{(0)\gamma}_{ll}(q^2) = {ie^2_0\over q^2}
\bar v_l\gamma_{\mu}u_l \bar u_l\gamma^{\mu}v_l  \eq

\bq    A^{(0)Z}_{ll}(q^2) = {i\over q^2-M^2_{0Z}}({g^2_0\over4c^2_0})
\bar v_l\gamma_{\mu}(g^{(0)}_{Vl}
-\gamma^5 g^{(0)}_{Al})u_l \bar u_l\gamma^{\mu}(g^{(0)}_{Vl}
-\gamma^5 g^{(0)}_{Al})v_l   \eq

\noindent
and (note the particular normalization)

\bq    A^{(0)Z'}_{ll}(q^2) = {i\over q^2-M^2_{0Z'}}({g^2_0\over4c^2_0})
\bar v_l\gamma_{\mu}(g^{'(0)}_{Vl}
-\gamma^5 g^{'(0)}_{Al})u_l \bar u_l\gamma^{\mu}(g^{'(0)}_{Vl}
-\gamma^5 g^{'(0)}_{Al})v_l    \eq

\noindent
($e_0\equiv g_0s_0$, $c^2_0\equiv 1-s^2_0$).\par
Following the usual approach, we shall treat the $Z'$ effect at one
loop in the SM sector and at "effective" tree level for the $Z'$
exchange diagram, whose interference with the analogous photon and $Z$
graphs will give the relevant virtual contributions. The $Z'$ width
will be considered "sufficiently" small with respect to $M_{Z'}$ to be
safely neglected in the $Z'$ propagator, and from what previously said
the $Z-Z'$ mixing angle will be ignored. If we stick ourselves to final
charged leptonic states, we must therefore deal with only two
"effective" parameters, more precisely the ratios of the
quantities $g'_{Vl}/\sqrt{M^2_{Z'}-q^2}$ and
$g'_{Al}/\sqrt{M^2_{Z'}-q^2}$, that contain the (conventionally
defined) "physical" $Z'$ mass and two "physical" $Z'll$ couplings,
whose meaningful definition would be related to a $Z'$ discovery and to
measurements of its various decays, that are obviously missing. This
will not represent a problem in our case since in our approach these
parameters, as well as any intrinsic overall (scale) ambiguity related
to their actual definition, will disappear in practice, being replaced
by model independent functional relationships between different
leptonic observables.\par
For what concerns the treatment of the SM sector, a prescription has
been very recently given \cite{Zsub} that corresponds to a "$Z$-peak
subtracted" representation of two-fermion production, in which a
modified Born approximation and "subtracted" one-loop corrections are
used. These corrections, that are "generalized" self-energies, i.e.
gauge-invariant combinations of self-energies, vertices and boxes, have
been called in refs.\cite{Zsub}, to whose notations and conventions we
shall stick, $\tilde{\Delta}\alpha(q^2)$, $R(q^2)$ and $V(q^2)$
respectively. As it has been shown in ref.\cite{Zsub}, they turn out to
be particularly useful whenever effects of new physics must be
calculated. In particular, the effect of a general $Z'$ can be treated
in this approach as a particular modification of purely "box" type to
the SM values of  $\tilde{\Delta}\alpha(q^2)$, $R(q^2)$ and $V(q^2)$
given by the following prescriptions:

\bq  \tilde{\Delta}^{(Z')}\alpha(q^2) =-
{q^2\over M^2_{Z'}-q^2}({1\over 4s^2_1c^2_1})g^2_{Vl}
(\xi_{Vl}-\xi_{Al})^2
\eq

\bq R^{(Z')}(q^2) = ({q^2-M^2_Z
\over M^2_{Z'}-q^2})
\xi_{Al}^2 \eq

\bq V^{(Z')}(q^2) = -
({q^2-M^2_{Z}\over M^2_{Z'}-q^2})({g_{Vl}\over2s_1c_1})
\xi_{Al}(\xi_{Vl}-\xi_{Al}) \eq

\noindent
where we have used the definitions :

\bq \xi_{Vl} \equiv {g'_{Vl}\over g_{Vl}}   \eq

\bq \xi_{Al} \equiv {g'_{Al}\over g_{Al}}   \eq

\noindent
with $g_{Vl}=-{1\over2}(1-4s^2_1)$; $g_{Al}=-{1\over2}$ and
$s^2_1\equiv1-c^2_1$; $s^2_1c^2_1=\pi\alpha(0)/\sqrt2 G_{\mu}M^2_Z$.\par
To understand the philosophy of our approach it is convenient to write
the expressions at one-loop of the \underline{three} independent
leptonic observables that will be measured at LEP2, i.e. the muon cross
section and forward-backward asymmetry and the final $\tau$
polarization (the latter quantity being theoretically equivalent to the
final lepton longitudinal polarization asymmetry, that might be
measured at a future $500$~GeV NLC). Leaving aside specific QED
corrections, these expressions read:

\bqa  \sigma_{\mu}(q^2)=&&\sigma^{Born}_{\mu}(q^2)\bigm\{1+{2\over
\kappa^2(q^2-M^2_Z)^2+q^4}[\kappa^2(q^2-M^2_Z)^2
\tilde{\Delta}\alpha(q^2)\nonumber\\
&&-q^4(R(q^2)+{1\over2}V(q^2))]\bigm\} \eqa

\newpage

\bqa  A_{FB,\mu}(q^2)=&&A^{Born}_{FB,\mu}(q^2)\bigm\{1+
{q^4-\kappa^2(q^2-M^2_Z)^2
\over\kappa^2(q^2-M^2_Z)^2+q^4}[
\tilde{\Delta}\alpha(q^2)+R(q^2)]\nonumber\\
&&+{q^4\over\kappa^2(q^2-M^2_Z)^2+q^4}V(q^2)]\bigm\} \eqa

\bqa A_{\tau}(q^2)=&&A^{Born}_{\tau}(q^2)
\bigm\{ 1+[{\kappa(q^2-M^2_Z)
\over\kappa(q^2-M^2_Z)+q^2}-{2\kappa^2(q^2-M^2_Z)^2\over
\kappa^2(q^2-M^2_Z)^2+q^4}]
[\tilde{\Delta}\alpha(q^2)\nonumber \\
&&+R(q^2)]
-{4c_1s_1\over v_1}V(q^2) \bigm\} \eqa

\noindent
where $\kappa^2$ is a numerical constant ($\kappa^2\equiv
({\alpha\over3\Gamma_l M_Z})^2\simeq 7$) and we defer to ref.\cite{Zsub}
for a more detailed derivation of the previous formulae.\par
A comparison of eqs.(10)-(12) with eqs.(5)-(7) shows that, in the three
leptonic observables, only \underline{two} effective parameters, that
could be taken for instance as $\xi_{Vl}M_Z/\sqrt{M^2_{Z'}-q^2}$,
$(\xi_{Vl}-\xi_{Al})M_Z/\sqrt{M^2_{Z'}-q^2}$ (to have dimensionless
quantities, other similar definitions would do equally well), enter.
This leads to the conclusion that it must be possible to find a
relationship between the \underline{relative} $Z'$ shifts
${\delta\sigma^{Z'}_{\mu}\over\sigma_{\mu}}$,
${\delta\A^{Z'}_{FB,\mu}\over\A_{FB,\mu}}$ and
${\delta\A^{Z'}_{\tau}\over\A_{\tau}}$ (defining the shift,
for each observable
$\equiv \O$, through $\O\equiv \O^{SM} + \delta \O^{Z'}$)
that is completely
independent of the values of these effective parameters. This will
correspond to a region in the 3-d space of the previous shifts that
will be \underline{fully characteristic} of a model with the
\underline{most general type} of $Z'$ that we have considered. We shall
call this region "$Z'$ reservation"\footnote{Reservation : "Tract of
land reserved for \underline{exclusive} occupation by native tribe",
Oxford Dictionary, 1950.}\par
To draw this reservation would be rather easy if one relied on a
calculation in which the $Z'$ effects are treated in first
approximation, i.e. only retaining the leading effects, and not taking
into account the QED (initial-state) radiation. After a rather
straightforward calculation one would then be led
to the following approximate expressions that we
only give for indicative purposes:

\bq  \bigm[{\delta A^{(Z')}_{\tau}\over  A_{\tau}}\bigm]^2  \simeq
f_1f_3({8c^2_1s^2_1\over v^2_1}){\delta\sigma^{(Z')}_{\mu}
\over \sigma_{\mu}}\bigm[{\delta A^{(Z')}_{FB,\mu}
\over  A_{FB,\mu}}+{1\over2}f_2{\delta\sigma^{(Z')}_{\mu}
\over \sigma_{\mu}} \bigm] \eq

\noindent
where

\bq  f_1 = {\kappa^2(q^2-M^2_Z)^2+q^4 \over \kappa^2(q^2-M^2_Z)^2} \eq

\bq   f_2 = {\kappa^2(q^2-M^2_Z)^2-q^4 \over \kappa^2(q^2-M^2_Z)^2} \eq

\bq   f_3 ={\kappa^2(q^2-M^2_Z)^2+q^4 \over \kappa^2(q^2-M^2_Z)^2-q^4}
 \eq

Eq.(13) is only an approximate one. A more realistic description can
only be obtained if the potentially dangerous QED effects are fully
accounted for. In order to accomplish this task,
the QED structure function formalism~\cite{qed} has been
employed as a reliable tool for the treatment of
large undetected initial-state photonic radiation. Using the
 structure function method amounts to writing, in analogy with
QCD factorization, the QED corrected cross section~\cite{prd} as
a convolution of the form~\footnote{\footnotesize The actual
implementation of QED corrections is performed,
in the Monte Carlo code, at the level of the
differential cross section, taking into account
all the relevant kinematical effects according to~\cite{prd};
in the present paper only
a simplified formula is described, for the sake of simplicity. }
\begin{eqnarray}
\sigma (q^2) = \int d x_1 \, d x_2
\, D(x_1,q^2) D(x_2,q^2) \sigma_0 \left( (1 - x_1 x_2) q^2 \right)
\left\{1 + \delta_{fs} \right\} \Theta({\rm cuts}),
\label{eq:master}
\end{eqnarray}
where $\sigma_0$ is the lowest-order kernel cross section, taken at the
 energy scale reduced by photon emission, and $D(x,q^2)$ is
the electron (positron) structure function.
Its expression, as obtained by solving the Lipatov--Altarelli--Parisi
evolution equation in the non-singlet approximation, is given
by~\cite{qed}:
\begin{eqnarray}
D(x,q^2)&=&\, \Delta^{'} \,
\hbeta (1 - x)^{\hbeta - 1} - {{\beta} \over 4} (1+x) \\ \nonumber
&+& {1 \over {32}} \beta^2 \left[ -4 (1+x) \ln(1-x) + 3 (1+x) \ln x -
4 {{\ln x} \over {1-x}} - 5 - x \right] ,
\end{eqnarray}
with
\begin{eqnarray}
\beta = 2\,{\alpha \over \pi}\,(L-1)
\end{eqnarray}
where $L= \ln \left( {q^2 / {m^2}} \right)$ is the collinear logarithm.
The first exponentiated term is associated to soft multiphoton emission,
 the second and third ones describe
single and double hard bremsstrahlung
in the collinear approximation. The $K$-factor $\Delta^{'}$ is of the form
\begin{eqnarray}
\Delta^{'} = 1 \,+\, \left(\alpha \over \pi \right) \Delta_1 \,+\,
\left(\alpha \over \pi \right)^2 \Delta_2
\end{eqnarray}
where $\Delta_1$ and $\Delta_2$ contain respectively
 $\cal O (\alpha)$ and $\cal O (\alpha^2)$ non-leading QED corrections
known from explicit perturbative calculations. The actual expression used for
these non-leading corrections is the one valid in the soft-photon
approximation, which is justified by the fact that, in order to avoid the $Z$
radiative return, a cut on the hard-photon tail is imposed.
 In eq.~(\ref{eq:master})
$\Theta({\rm cuts})$ represents the rejection algorithm
to implement possible experimental cuts, $\delta_{fs}$ is the correction
factor to account for QED final-state radiation. Since only a
cut on the invariant mass $s^{'} =
x_1 x_2 q^2$ of the event after initial-state radiation is imposed
in our numerical analysis (see below), the simple formula
$\delta_{fs} = 3 \pi / 4 \alpha $ holds.
In order to proceed with
the numerical simulation of the
$Z^{'}$ effects under
realistic experimental conditions, the master formula~(\ref{eq:master})
has been implemented in a Monte Carlo event generator which has been
first checked against currently used LEP1 software~\cite{topaz0},
found to be
in very good agreement and then used to produce our numerical
results. The $Z^{'}$ contribution has been included in the
kernel cross section $\sigma_0$ computing the
$s-$channel Feynman diagrams associated
 to the production of a $l {\bar l}$ pair in a
$e^+ e^-$ annihilation mediated by the exchange of a photon, a
standard model $Z$ and an additional $Z^{'}$ boson. In the calculation,
which has been carried out within the helicity amplitude
formalism for massless fermions and with
the help of the program for the algebraic manipulations
{\tt SCHOONSCHIP}~\cite{schoon}, the coupling of the $Z^{'}$
 boson to the leptons has been parametrized, as already
pointed out, as:
\begin{eqnarray}
\gamma^{\mu} \, \left( g^{'(0)}_{Vl} \,-\, g^{'(0)}_{Al}
\gamma_5 \right)
\end{eqnarray}
and the $Z^{'}$ propagator has been included in the zero-width
approximation (see above). Moreover, the bulk of non-QED corrections
has been included in the form of Improved Born Approximation,
choosing ${\bar \alpha}(q^2), M_Z, G_F$, together with $\Gamma_Z$,
 as input parameters.
The values used for the numerical simulation are~\cite{pdg}:
$M_Z = 91.1887$~GeV, $\Gamma_Z = 2.4979$~GeV; the center of mass
energy has been fixed at a typical LEP2 value $\sqrt{q^2} = 175$~GeV and
the cut $s^{'} / q^2 > 0.35$ has been imposed in order to
remove the events due to $Z$ radiative return and hence disentangle
the interesting virtual $Z^{'}$ effects. These have been
investigated allowing the previously defined ratios
$\xi_V$ and $\xi_A$ to vary within the ranges $-2 \leq \xi_A \leq 2$ and
$-10 \leq \xi_V \leq 10$. Higher values might be also taken
into account; the
reason why we chose the previous ranges was that, to our knowledge, they
already include all the most popular existing models.

The results of our calculation are shown in Fig.~1. The central box
corresponds to the ''dead'' area
where a signal would not be distinguishable corresponds to an assumed
(relative) experimental error of 1.5\% for $\sigma_\mu$ and to 1\%
(absolute)
errors on the two asymmetries. The region that remains outside
the dead area
represents the $Z'$ reservation at LEP2, to which the effect of the most
general $Z'$ must belong.

One might be interested in knowing how different the realistic
Fig.~1 is from
the approximate ''Born'' one, corresponding in particular to
the simplest
version given by eq.~(13). This can be seen in Fig.~2, where
we showed the
two surfaces (the points correspond to the realistic
situation, Fig.~1). One
sees that the simplest Born calculation is, qualitatively,
a good approximation
to a realistic estimate, which could be very useful if one
first wanted to look
for sizeable effects.

The next relevant question that should be now answered is whether the
correspondence between $Z'$ and reservation is of the one to
one type, which
would lead to a unique identification of the effect. We
have tried to answer
this question for one specific and relevant case, that of
virtual effects
produced by anomalous gauge couplings~(AGC). In particular,
we have considered
the case of the most general, dimension six, $SU(2)
\otimes U(1) $ invariant
effective Lagrangian recently proposed~\cite{Hagi}. This has been fully
discussed in a separate paper~\cite{RV}, where the previously mentioned
''$Z$-peak subtracted'' approach has been used. The resulting
AGC reservation
in the ($\sigma_\mu$, $A_{FB,\mu}$, $A_{\tau}$) plane is a
certain region,
drawn, for simplicity, in Born approximation as suggested by
the previous
remarks. In Fig.~3 we
have plotted this region and it can be compared to the general
$Z'$ one plotted in Fig.2. As one
sees, the two reservations do not overlap in the meaningful
region. Although we
cannot prove this property in general, we can at least conclude
that, should a
clear virtual effect show up at LEP2, it would be possible to decide
unambiguously to which among two very popular proposed models
it does belong.

If the signal belonged to the $Z'$ reservation, the immediate
request would be
to identify its origin. Clearly, this would imply a knowledge
of the $Z'$
mass, that could only be achieved by future direct production,
but this
discussion is clearly beyond the purposes of this paper.
The point that we
wanted to raise here is that LEP2 might already provide
convincing indications
of the existence of a $Z'$ \underline{before} it is
actually discovered. This
would generalize to the New Physics sector the previous
remarkable prediction
of LEP1 for the top quark.

\newpage

\centerline { {\bf Figure Captions }}

\vspace{0.5cm}

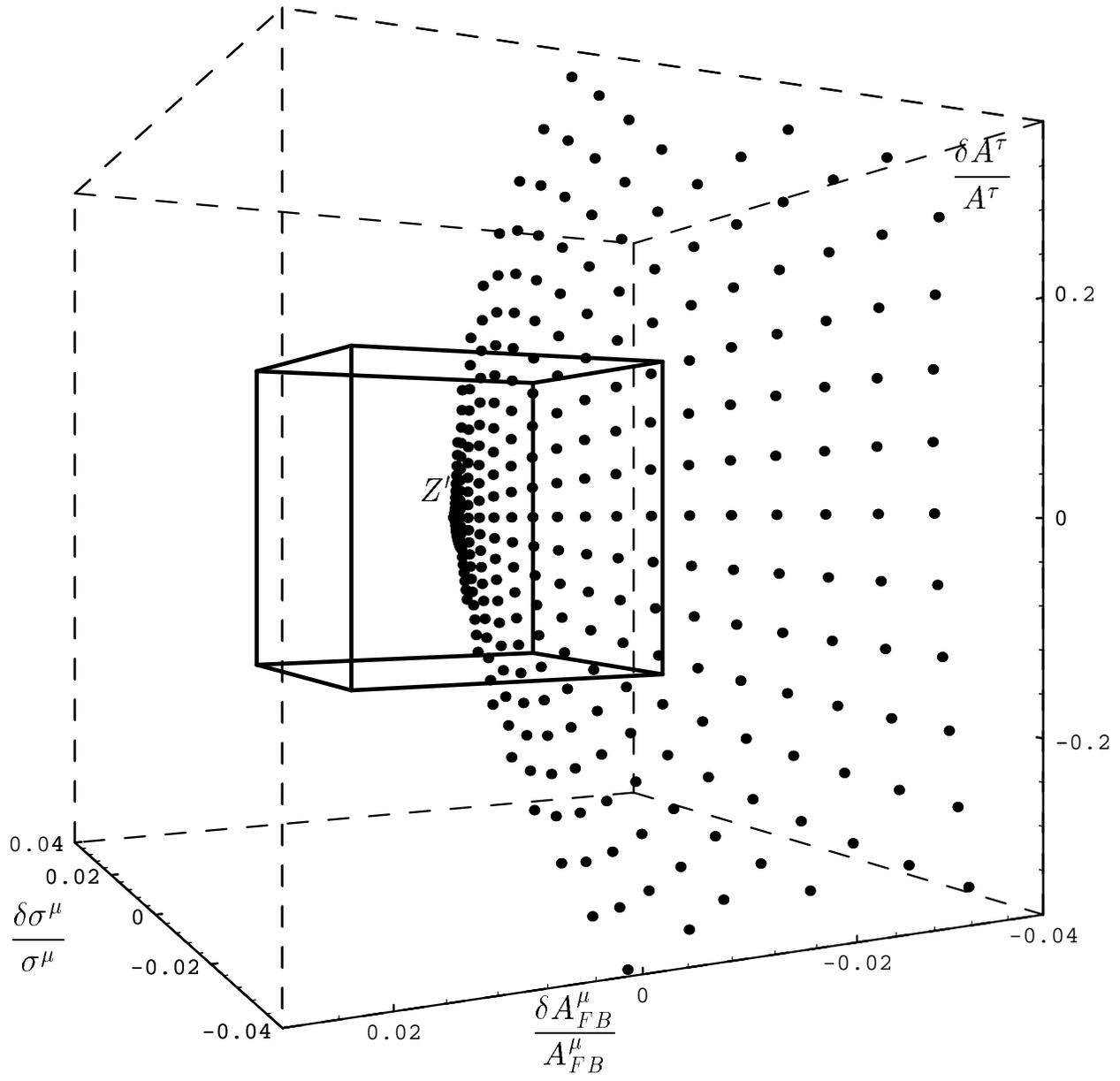
{\bf Fig.~1} {${{\delta A^\tau} \over {A^\tau}} $ versus
${{\delta \sigma^\mu} \over {\sigma^\mu}} $  and
${{\delta A^\mu_{FB}} \over {A^\mu_{FB}}} $.
The central ''dead'' area
where a signal would not be distinguishable corresponds to an assumed
(relative) experimental error of 1.5\% for $\sigma_\mu$ and to
1\% (absolute)
errors on the two asymmetries. The region that remains outside
the dead area
represents the $Z'$ reservation at LEP2, to which the effect of
the most
general $Z'$ must belong.}

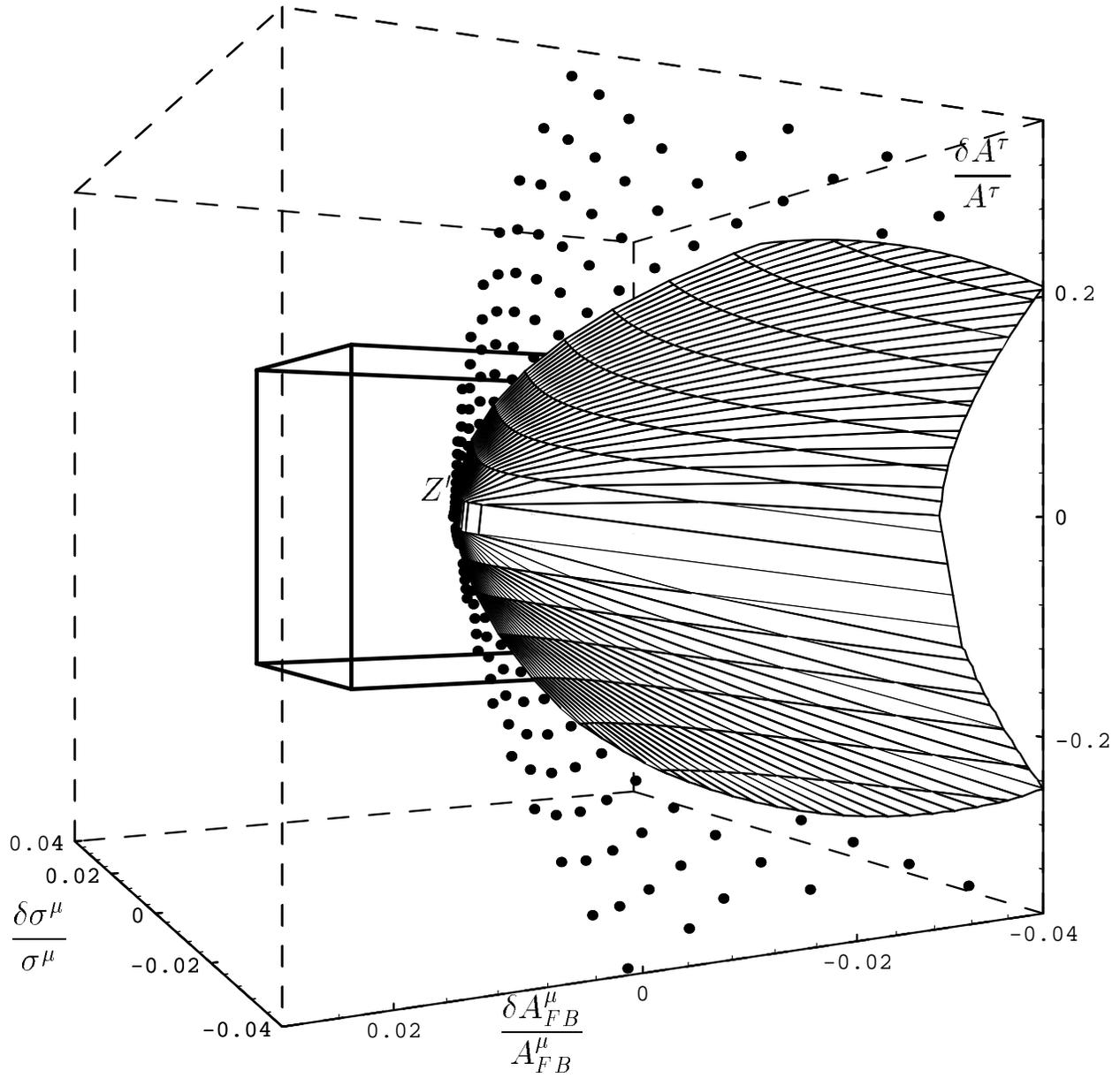
{\bf Fig.~2} {The same as Fig.~1, comparing the realistic results
obtained via
Monte Carlo simulation with the approximate ones according to
eq.~(13). }

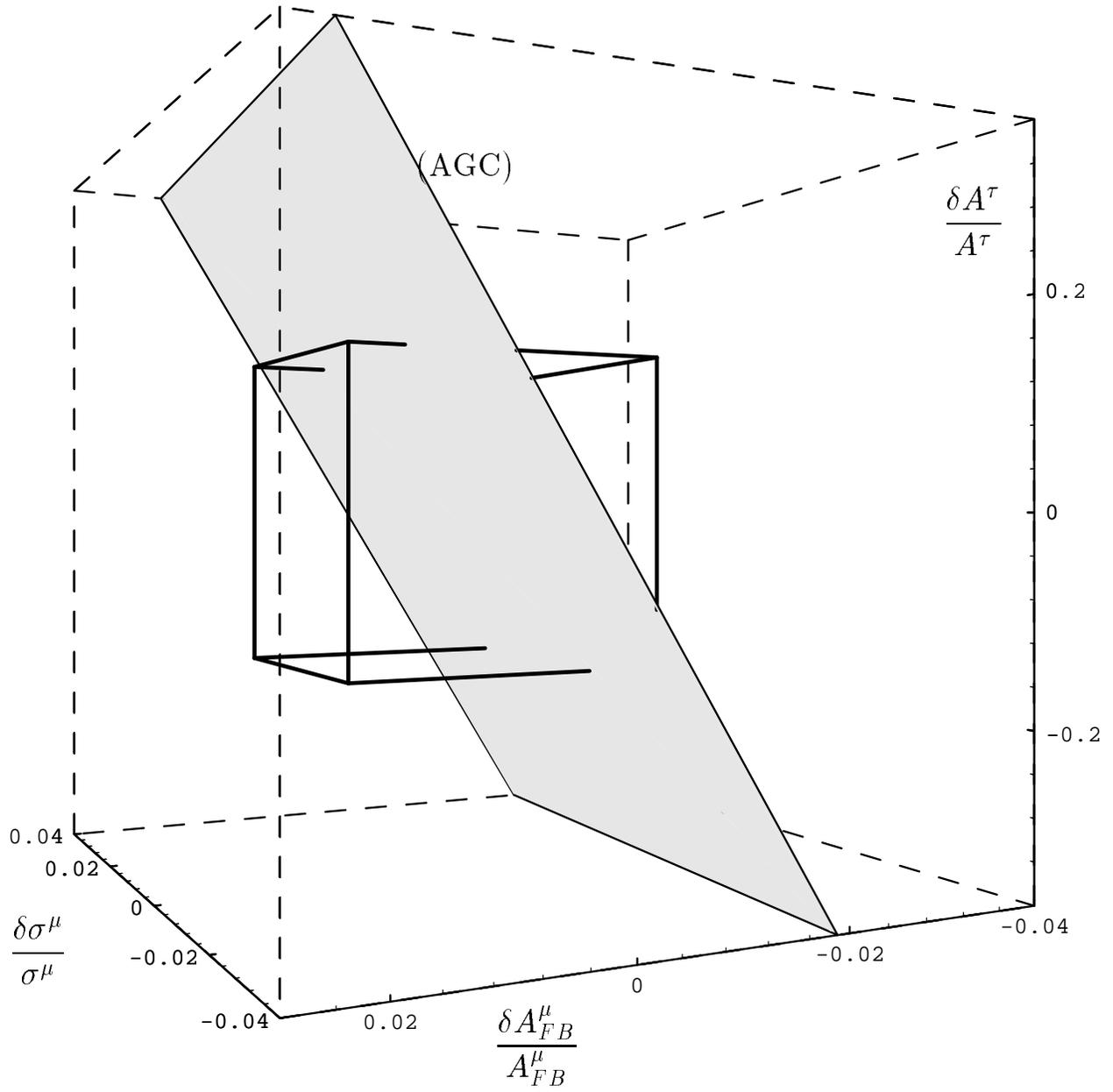
{\bf Fig.~3} {The same as Fig.~2, showing  the region
corresponding to AGC. }

\end{document}